# Cavitation of matter in the Universe into a degenerated Voronoi foam

T. Nørretranders[1], J. Bohr[2], and S. Brunak[3]

[1]Onsgårdsvej 19, DK-2900 Hellerup, Denmark.
[2]Department of Solid State Physics, Risø National Laboratory,
 DK-4000 Roskilde, Denmark.
[3]Department of Physical Chemistry, The Technical University of Denmark,
 DK-2800 Lyngby, Denmark.

Fluids suspended in expanding laboratory systems cavitate into Voronoi foams that degenerate through secondary cavitation of the foam walls. This robust morphological mechanism results from a virtual competition for space between the cavities. In cosmology, Voronoi foams are known to explain important features of the large scale structure, but the physical mechanism for the formation of such foams is unknown. We here present a scenario for the formation of large scale structure: Cavitation initiates at the abrupt reduction of radiation pressure on matter at decoupling and results in the cold flow of matter into a single, void-dominated structure with a scale determined by the Hubble sphere at decoupling. This mechanism does not impose restrictions on the smoothness of the last scattering surface at decoupling and conserves the entropy of matter during the self-organization of a single cavitation foam spanning the observable universe.

# 1 Introduction

Understanding the formation of large scale structure in the universe has become a challenge for the standard model of cosmology (Peebles et al. 1991). The high degree of isotropy of the cosmic background radiation indicates that the distribution of matter in the early universe was smooth (Smoot et al. 1991). However, in the presently observed universe, the distribution of luminous matter is highly inhomogenous, on the large scale displaying a void-dominated structure of sheets and filaments (Zeldovich et al. 1980; De Lapparent et al. 1986; Bachall 1988; Geller & Huchra 1989; Saunders et al. 1991). Models employing a host of physical processes, many of them involving novel elementary particles or gravitational phenomena, have failed to reproduce the observed structure from the minor density fluctuations left over from the early universe (Peebles & Silk 1990).

In this report we present a scenario for the formation of structure where density perturbations are translated into a Voronoi foam structure through a morphological mechanism observable in the laboratory. Whereas earlier interpretations of the large scale structure as a Voronoi foam are primarily based on the geometrical similarity, our scenario is rooted in a physical mechanism reproducible in the laboratory and applicable to the cosmological problem.

# 2 Laboratory fluids in expanding systems

In a study of the formation of viscous trees (Bohr et al. 1993), we observed the formation of a cellular structure akin to a Voronoi tessellation (Voronoi 1908; Møller 1989) when a viscous fluid was sandwiched between two parallel plates which subsequently were separated, see Fig. 1. The time evolution of the process displays the growth of cavities and the organization of the fluid into smooth membranes surrounding voids.

## 2.1 Primary cavitation

The formation of cavities in the fluid is due to hydrodynamic cavitation, the formation of cavities by pressure reduction (Young 1989), in this case resulting from the suspension of a non-expandable fluid in an expanding system. Since the fluid cannot expand, cavities appear and grow as the the system expands. The morphological appearance of the process is akin to boiling, although no vapour pressure is needed to drive the cavitation since the hydrodynamic pressure of the fluid would become negative if it did not cavitate.

## 2.2 Smooth membranes

Although the mechanism is driven by forces acting upon the fluid, it can be expressed as a virtual competition for space between the cavities, leading to the formation of smooth membranes in the fluid.

Voronoi foams arise during the separation of plates (see Fig. 2) suspended by a viscous fluid when the ambient pressure is negligible compared to the negative hydrodynamic pressure in the fluid, e.g. under vacuum or with high pulling speed. When an ambient pressure is quantitatively efficient, viscous trees (see Fig. 3) arise, also involving the formation of smooth membranes. The ubiquitous formation of viscous trees at ambient pressure involves an *actual* competition for space between viscous fingers penetrating the periphery of a viscous fluid, resulting in the formation of smooth membranes between the fingers. The formation of the Voronoi foam, however, can take place under vacuum inside a fluid with negligible vapour pressure, i.e. in the absence of any agent mediating an actual competition for space. The similarity between the morphological mechanisms of actual and virtual competition for space is due to the self-organization of the fluid into one membrane structure spanning the expanding system.

## 2.3 Voronoi morphologies

The Voronoi foam results from a morphological mechanism that allows the fluid to conserve its volume while spanning the expanding system. A Voronoi tessellation is defined by a discrete set of seed points that partitions the plane into cells containing the part of the plane that is closer to the cell seed point than to any other seed point; a Voronoi foam is the corresponding partition of space into polyhedra. In the intermediate stage, where the fluid in the expanding system is not yet cavity-dominated, the fluid represents a dynamic Johnson-Mehl tessellation (Johnson & Mehl 1939; Zaninetti 1989) where cavities grow and compete for space in the plane of the plates. In our experimental system, where the fluid is suspended between two parallel plates, the system is three-dimensional and the structure consequently a foam. However, the structure is essentially confined to two dimensions and, when viewed along the axis of plate separation, resembles a tessellation. The Voronoi pattern represents the positions where the fluid has only moved in the direction of expansion during the cavity formation process; hence, the footprint of the foam is a tessellation. The seed points of the foam are distributed at locations where the tensile strength of the fluid first becomes less than the tension due to the negative hydrodynamic pressure.

## 2.4 Secondary cavitation

At later stages in the development of the foam, a secondary cavitation takes place. As the membranes get thinner, they can no longer span the expanding system. They therefore cavitate, displaying growing holes, see Fig. 4. This secondary cavitation takes place inside the individual sheets of the foam and reduces them to the strings of the foam edges; a corresponding tertiary cavitation, in which the edges of the foam cells break and are reduced to the vertices, is responsible for the cleavage of the foam into the two mirror-image tessellations left behind on the surfaces of the plates. The two-dimensional laboratory foam structure has an experimentally determined (computed by the box counting method) fractal dimension of 2.0 when considered on the largest length scales.

We suggest that this cavitation process in fluids is analogous to the one that led to the origin of large scale structure in the Universe: fluctuations left over from the early universe were translated into a Voronoi foam by a cavitation process initiated at the event of the abrupt reduction of radiation pressure at decoupling. Such a cavitation process results in a distinct structure regardless of the nature and density of the fluctuations and is thus compatible with the spectrum of fluctuations predicted by inflation (Kashlinsky & Jones 1991) and witnessed by the COBE experiment (Smoot et al. 1992).

A common two-dimensional model of the Hubble expansion is the surface of an inflated balloon (Eddington 1930). Cosmic cavitation can be modelled as a balloon covered by a two-dimensional layer of a non-expandable fluid and then inflated; since the fluid cannot adapt to the larger area by thinning, it must cavitate and form voids that grow with the expansion of the balloon.

# 3   Voronoi foams in cosmology

Voronoi foams have recently been employed with much success to describe the large scale structure in the universe. In 1984, Matsuda and Shima suggested that the shape of superclusters are akin to the convex polyhedra of a spatial Voronoi tessellation (Matsuda & Shima 1984). In 1988, van de Weygaert and Icke found that the two-point correlation function of the Voronoi foam vertices agrees with the observed distribution of Abell clusters (van de Weygaert & Icke 1989; Icke & van de Weygaert 1991; Yoshioka & Ikeuchi 1989). Later, the multifractal spectrum of the CfA redshift survey of galaxies was reproduced by a Voronoi foam which has the correct correlation and Hausdorff dimension ($D_2 = 1.4$ and $D_H = 2.0$) (Martinez et al. 1990); the geometry of Voronoi foams explained (Coles 1990; Coles 1991; van de Weygaert 1991; Ikeuchi & Turner 1991) the apparent periodicity found in deep pencil beam surveys of galaxies at high redshifts (Broadhurst et al. 1990).

However, the particular physical models previously suggested for the formation of such foams have proved inadequate in explaining essential features of the observed universe. The kinematic model, based on the expansion of underdensities and the subsequent streaming of matter into a Voronoi foam (Icke & van de Weygaert 1991), cannot explain the isotropy of the background radiation (Coles & Barrow 1990); explosion models (Ostriker & McKee 1988) cannot explain voids of the observed size (Ostriker & Strassler 1989; Weinberg & Ostriker 1989); Voronoi models based on cosmic strings (Zaninetti 1989) and global textures (Turok 1991) involve mechanisms which at present lack empirical foundation. Further, the scale length of the galaxy-galaxy correlation function (Williams et al. 1991) and genus measures of the topology of the observed structure (Moore et al. 1992) are not accounted for in the existing cellular models.

It is therefore of interest to explore an alternative physical mechanism leading to the formation of a Voronoi foam.

# 4 Matter in the Universe treated as a viscous fluid

In order to explore the analogy to viscous fluids in expanding laboratory systems, we treat matter in the post-decoupling universe as if it behaved like a viscous fluid in the non-expandable limit. Thus the matter-fluid is treated as one single body. We thus demand that matter in the Universe exhibits an internal attractive interaction that simulates connectivity and viscosity; that the entropy is conserved and that there is a tension mediating the negative hydrodynamic pressure from the expansion.

## 4.1 Cavitation

During all epochs, radiation behaves as a relativistic gas that remains in equilibrium with the expansion through stretching of the wavelength proportionally to the cosmic scale factor (Harrison 1993). However, matter behaved differently before and after the decoupling of matter and radiation. Prior to the decoupling, the radiation pressure dampened all fluctuations and kept matter in equilibrium with the expansion. At the advent of the hydrogen atom, matter decoupled and no longer participated in the expansion; the scale of the atom is governed by quantum mechanics and does not change.

It then became energetically unfavourable for matter to undergo uniform expansion. Cavitation became favourable, since a local attractive interaction in the matter-fluid with a potential decreasing with distance less rapidly than linearly (the gravitational potential is inversely linear) allows instabilities to develop into cavities. The gravitational potential acts analogously to the attractive term of the Lennard-Jones potential in a laboratory fluid.

## 4.2 Entropy conservation

As radiation was stretched with the scale factor, matter cavitated into a foam structure with a characteristic scale that varies linearly with the scale factor. In both cases, the entropy remains constant during the adiabatic expansion. Radiation entropy is known to be conserved because wave stretching reduces the temperature as the volume expands (Kolb & Turner 1990). Matter entropy is however also conserved during cavitation since the addition and expansion of voids does not increase the number of degrees of freedom available to matter organized into a foam. The intimate relationship between algorithmic information content (Chaitin 1987) and physical entropy (Zurek 1989a,b) allows the cavitation process to be modelled by a random string of binary digits in which a string of zeroes is inserted in one place. Since the algorithmic information content $K(s)$ of a string $s$ is given by the size in bits of the shortest program on a universal computer that will reproduce $s$, the information content of a random string is equal to its length. However, an ordered string, such as a string of $N$ zeroes, can be specified by only $\log_2 N + O(1)$ bits, where the term $O(1)$ specifies the address of insertion and the value of the inserted ordered string, while the logarithmic term specifies its length. Thus $K(s)$ of a random string does not grow more than $\log_2 N + O(1)$ bits when a string of $N$ zeroes is inserted in one place (and only $\log_2 \Delta N$ bits when the string grows). Since physical entropy is equal to $K(s)$ for an observer having full information of the microstate (Zurek 1989 a,b), the shortest possible description of the distribution of matter does not increase with more than $\log_2 N + O(1)$ bits with each cavity of size $N$ introduced. Thus the entropy of the expanding universe, except for a logarithmic correction, remains constant due to the stretching of radiation and the cavitation of matter.

## 4.3 Scale of cavities and void sizes

In the following we interpret such cavities as the voids reported in redshift surveys. Due to cavitation, voids appeared in the matter-fluid and soon grew proportional to the expansion

$$L_v(t) \propto (R^3(t) - R^3(t_*))^{1/3} \simeq R(t) \text{ for } R(t) >> R(t_*), \tag{1}$$

where $L_v(t)$ is the void size, $R(t)$ is the cosmological scale factor and $R(t_*)$ is the scale at decoupling. Voids presently of size $L_v(0)$ had the size $L_v(z) = L_v(0)/(1+z)$ at earlier epochs, except for the initial cavitation at decoupling when void size must be described by volume conservation of non-expanding matter in an expanding universe,

$$L_v(z) = L_v(0) \left( \frac{R^3(z) - R^3(z_*)}{R^3(0) - R^3(z_*)} \right)^{1/3} \simeq \frac{L_v(0)}{1+z}, \tag{2}$$

where $z_*$ is the redshift of decoupling.

Voids observed today have a typical size of $50 - 100 h^{-1}$ Mpc (Blumenthal et al. 1992). Assuming linear growth $L_v \propto R(t)$, the upper limit to the void size at decoupling is $0.05 - 0.1 h^{-1}$ Mpc, well below the particle horizon at decoupling, $L_P = 0.2 h^{-1}$ Mpc. The particle horizon is the observable universe (all world lines that in principle can be observed) of the epoch. Thus the actual void size in the decoupling era diverged from zero to a value compatible with the horizon size.

## 4.4 Hubble sphere tension

The Hubble sphere (Harrison 1991) contains all astronomical systems that recede from the observer with a Hubble expansion velocity less than the velocity of light, $c$,

$$L_H = c/H, \tag{3}$$

where $L_H$ is the Hubble sphere radius and $H$ the Hubble parameter of the epoch.

In a matter-dominated Einstein-de Sitter universe with flat space ($\Omega = 1$), $L_P = 2L_H$, $H = \frac{2}{3}t^{-1}$ and $H(z) = H_0(1+z)^{\frac{3}{2}}$ (Harrison 1993). Thus at decoupling ($z_* = 1000, t = 3 \times 10^5$ yr), $H$ was $3 \times 10^6 h^{-1}$ km/s/Mpc (i.e. $\sim 10c$/Mpc) and $L_H = 0.1 h^{-1}$ Mpc.

The physical mechanism mediating a negative hydrodynamic pressure on the matter-fluid is an internal tension in a body due to the Hubble expansion. This tension becomes effective when the radiation pressure on matter is reduced at decoupling. A fluid spanning over the entire Hubble sphere must cavitate with at least one cavitation point per Hubble sphere. This cavitation is the result of the tension inside an object extending over the velocity dispersion $\Delta v$ approaching $c$. As the size $l$ of a single, locally connected object approaches $L_H$ the internal tension will diverge since the kinetic energy per length unit will diverge. This is due to the relativistic increase in mass resulting from $\Delta v = lH \sim c$ for $l \sim L_H$.

After the abrupt drop in radiation pressure on matter at decoupling, the matter-fluid will experience this Hubble sphere tension and consequently cavitate.

The Hubble sphere thus defines the maximum distance between the cavitation points. The minimum distance is determined by the effective tensile strength of the matter-fluid immediately after decoupling. Since the internal tension diverges as the size of the object approaches the Hubble sphere, it is likely that the actual distance beween the cavitation points is of the same order of magnitude as the Hubble sphere.

After the decoupling, the cavities grow with $[R(t)^3 - R(t_*)^3]^{\frac{1}{3}} \sim R(t)$ and almost span the entire volume, interspaced only by the non-expanded walls of the matter-fluid. Thus the uniformly distributed matter-fluid is transformed into a single foam-structure of cavities surrounded by sheets. The distance between the cavitation points $d_C$ for $z < z_*$ is

$$d_C(z) = d_C(z_*) \frac{1 + z_*}{1 + z}, \qquad (4)$$

with $d_C(z_*) < L_H(z_*) = 0.1 h^{-1}$ Mpc. At present, $d_C < 100 h^{-1}$ Mpc and the walls of non-expanding matter are of the order of $0.1 h^{-1}$ Mpc. The density of cavitation points, expressed by $d_C(z)$, represents the only scale parameter in the Voronoi foam structure. The assumption of a Poissonian distribution makes $d_C$ the only free parameter. However, since the local release of Hubble sphere tension resulting from cavitation will reduce the local probability of the formation of yet a cavity, the distribution is likely to be less random than a Poissonian distribution.

## 5 Comparison with observed features

### 5.1 Scale of structure

An average seed point separation of $\sim 50 - 100 h^{-1}$ Mpc is not incompatible with the existing knowledge of the large scale structure. The observationally determined spacing between Voronoi foam walls at the present epoch (in which the walls are negligible and the void diameters are of the order of $d_C$) is $40 - 100h^{-1}$ Mpc, with the lower value of $40h^{-1}$ Mpc stemming from void sizes in surveys of the local structure (Blumenthal et al. 1992; Slezak et al. 1993) and the higher value of $100h^{-1}$ Mpc from the scale of the cluster correlation function (van de Weygaert & Icke 1989; Yoshioka & Ikeuchi 1989) and interpretations of deep surveys (Broadhurst et al. 1990). The interpretation of foam vertices as the Abell clusters gives a seed point spacing of $100h^{-1}$ Mpc (van de Weygaert & Icke 1989; van de Weygaert 1991). In the calibration of the Voronoi foam cell size to the periodicity found in deep pencil beam surveys, the mean spacing of walls for a foam with seed point separation of $100h^{-1}$ Mpc, originally estimated to be $137h^{-1}$ Mpc (Coles 1990), is $69h^{-1}$ Mpc (Coles 1991). This lower value may explain the apparent periodicity of $128h^{-1}$ Mpc in deep surveys when the beam size of the pencil beam is taken into account (Ramella et al. 1992): since a narrow beam does not hit a galaxy or cluster in every wall, the pencil survey will fail to detect all intervening structure and hence overestimate the size of the voids. A wall spacing of $\sim 69h^{-1}$ Mpc is also in accordance with the CfA survey void size of $50h^{-1}$ Mpc (Slezak et al. 1993). The galaxy-galaxy correlation function is not explained by this scaling of a random Voronoi foam (Williams et al. 1991).

## 5.2 Isotropy of cosmic background radiation

Since it results in a Voronoi foam of the correct scale, the cavitation model can reproduce the correlation function and the periodicity of red shift surveys, as can the kinematic model. However, unlike the kinematic model of expanding underdensities (Coles & Barrow 1990), the cavitation scenario does not impose any restrictions on the smoothness of the last scattering surface at decoupling. Any spectrum of fluctuations will be reduced by the cavitation mechanism into a Voronoi foam structure. After last scattering the voids will not influence the isotropy of the background radiation at a measurable level (Thompson & Vishniac 1987).

## 5.3 Increment in number of observable seed points

The number of seed points inside the horizon is presently much larger than at decoupling since $L_V \propto d_C \propto R(t) \propto t^{\frac{2}{3}}$ while $L_H \propto L_P \propto t$. The number of new cavitation points that become visible every year, $\Delta N_C$, is given by a shell around the sphere defined by $z = z_*$. The thickness of the shell is given by the rate of recession of the particle horizon, $dL_P/dt = 3c$ (Harrison 1991), and its radius by $L_P - ct_* \sim 5.9h^{-1}$ Gpc. With $d_C(z_*) = 0.1h^{-1}$ Mpc, $\Delta N_C = 10^5$. In principle, the energy released by the initial cavitation process can reach diverging values, since the velocity dispersion diverges at the birth of a cavity. We note that presently a full sky rate of $\sim 800$ short bursts of gamma-rays is detected every year (Higdon & Lingenfelter 1990). Their distribution is isotropic but not homogenous (i.e. a limited sphere centered on the observer). The luminosity of gamma-ray bursts, if situated at cosmological distances, is extremely high: $10^{53}$ erg released in ~1 second (Paczynski 1986).

## 5.4 Topology of structure

The application of a genus measure of topology to the IRAS redshift survey was found to be inconsistent with a Voronoi foam model (Moore et al. 1992), since the galaxy distribution was found to be more sponge-like than a cellular structure. However, the cavitation process not only leads to the Voronoi foam as a primary result, but also to secondary cavitation resulting in the opening of foam walls and consequent merging of voids. The topology of the structure changes during expansion. Before cavitation, the genus (defined as the number of holes minus the number of isolated regions) is $-1$ since matter is treated as a single fluid; after primary cavitation, the genus becomes very large and positive (one fluid with many holes); after secondary cavitation, the value decreases while remaining positive (one fluid with coalescing holes); after tertiary cavitation the cavitation foam degenerates into the vertex points and the genus becomes highly negative (one hole with many isolated regions). In other words, the cavitation process progressively changes a sphere into a foam with coalescing voids that eventually degenerates into isolated regions. The degeneration of the cavitation foam thus suggests a possible explanation for the observed topology. Also, the number of faces per polyhedron in the original foam is highly sensitive to the distribution of seed points.

## 5.5 Quasar Lyman-$\alpha$ forests

The deepest probe of structure in the Universe that is presently feasible is the study of the Lyman-$\alpha$ forests of quasars (Oort 1981) where the spectrum of the highly redshifted object displays absorption lines from matter with less redshift. Studies of the distribution of absorption lines have shown clustering of lines towards high redshifts (Peterson 1978),

$$dN/dz \propto (1+z)^y, \tag{5}$$

where $dN/dz$ is the number of absorption lines per redshift interval and $y$ is observationally determined to be around 2.5 (Weymann et al. 1981; Fabian & Barcons 1991). This clustering is due to an evolution effect (Pierre et al. 1988; Hoell & Priester 1991) most readily explained by a lower void filling factor $f_v$ at high redshifts (Pierre 1990), as a result of the growth of voids during time. However, $L_v \propto R(t)$ does not quantitatively reproduce the exponent of the clustering power law. A comoving population of fixed-size absorbers would (at $\Omega = 1$) give an exponent of 0.5 (Peterson 1978; Sargent et al. 1980), deviating strongly from the observed value. If the absorbers reside in the foam cell walls, not only does the wall separation of the foam grow with $R(t)$, the surface of the essentially two-dimensional cell walls will also grow with $(R(t))^2$. If the wall undergoes secondary cavitation (or if the matter inside the wall is unevenly distributed into discrete structures such as galaxies), the probability that the essentially one-dimensional light beam from the background quasar encounters a concentration of matter will decrease as the cell surface grows. Thus a factor of up to 2 may be added to $y$, resulting in a value for $y$ of 2.5, which could provide an explanation for the observed value. The observed coldness of the absorbing matter (Pettini et al. 1990; Stanek 1993) is also a natural consequence of cavitation.

## 5.6 Scale of galaxies

The characteristic scale of the walls and filaments of the foam is $\sim d_C(z_*) \leq 0.1 h^{-1}$ Mpc, well in accordance with the scale of galaxies. The scale of the foam vertices, where superclusters form, is several times larger. The foam structure forms a natural habitat for the dark halo matter known from studies of their dynamics to surround galaxies and clusters (Rubin 1983).

## 5.7 Cold flow of matter

The scenario predicts a very cold flow of matter, except close to the event of the formation of cavities. The velocity dispersion at the void-matter interface is initially very high, but soon becomes very low as the voids grow. For a model with a single void, the velocity dispersion at the void-fluid interface is given by

$$\Delta v = ((z_* + 1)^6/((z_* + 1)^6 - 2(z + 1)^3(z_* + 1)^3 + (z + 1)^6))^{1/3} - 1. \qquad (6)$$

For $z < 550$, $\Delta v < 0.1$. Thus at low $z$, the velocity field approaches the mean expansion.

The cold flow reflects the property of the cavitation process that matter does not move globally while the universe expands: matter is diluted by vacuum organized into voids. Apart from possible local turbulence at the matter-void interface, matter experiences only the growing gravitational potential arising from the expansion of voids.

Inside the individual sheets of the foam, local movement towards the nearest vertex due to a secondary cavitation of the foam sheets could account for peculiar motions such as the infall of the Local Group towards the Virgo cluster. Vertex-vertex peculiar motions could occur as a consequence of secondary and tertiary cavitation; this is qualitatively in accordance with the finding of strong intra-cluster motions (Bertschinger et al. 1991).

## 5.8 Non-emptiness of voids

In the laboratory experiments at low pulling speeds, the void-fluid interface becomes unstable with respect to Saffman-Taylor instabilities (Saffmann & Taylor 1958) and the circumference of the cavities displays viscous fingering (Bohr et al. 1993) of the same nature as that observed in the axisymmetric Hele-Shaw cell, where the less viscous fluid is injected at the centre of the more viscous fluid (Homsy 1987). This fingering results in tip-splitting and the consequent formation of minor filaments of fluid peninsulating the cavities. In cosmic cavitation, a similar process could account for the non-emptiness of the observed voids (Dey et al. 1990).

# 6  Conclusion

The advent of the hydrogen atom at decoupling led to the formation of a non-expandable fluid that underwent cavitation while the radiation was stretched by expansion. The cavitation led to a Voronoi foam that is in agreement with the observed large scale structure.

The scale of the foam is determined by the Hubble sphere at decoupling and the resulting morphology is consistent with the observed correlation function and multifractal spectrum of galaxy clusters, the periodicity in pencil-beam surveys, the topology of the IRAS survey, the evolution effect in quasar Lyman-$\alpha$ forests, the cold flow of matter and the conservation of the entropy of matter during the Hubble expansion. The morphology also suggests possible explanations for large scale motions and cosmic gamma ray bursts.

There is no physical reason for such a cavitation foam to be limited to the observed universe. It is likely that it encompasses the entire visible universe and reaches further.

*Acknowledgements.* We thank Drs. P. Kjærgaard Rasmussen, B. Gustafsson and M. Rees for comments and advice on the manuscript. Carsten Langemark kindly supplied the object depicted in Fig. 3. TN thanks the Royal Danish Academy of Fine Arts and Kulturfonden for support.

# Figure captions

Fig 1: Voronoi foam emerging following separation of two parallel plates in between which a fluid has been sandwiched. The material used was high vacuum silicone grease (Wacker–Chemie) with a viscosity of $5 \times 10^3$ Pa·s at a velocity gradient of $0.1$ s$^{-1}$. The experiment was performed in an evacuated container (pressure 100 Pa, substantially above the vapour pressure of the grease, 0.02 Pa) with a constant pulling force of 200N. One of the separated plates was transparent to allow visual inspection of the footprint of the fluid. In *b)* the fluid has been redistributed into a pattern akin to a Voronoi tessellation. In *a)* the intermediate stage of the cavitation process is shown. The duration of the process is of the order of seconds. Similar foams can be obtained at ambient pressure by increasing the pulling speed.

Fig 2: Manual production of viscous trees and Voronoi tessellations. The fluid is deposited on one of the two plates and then sandwiched between the plates. The early separation of the plates results in the formation of a menicus-dominated column of fluid. After further separation, the resulting viscous tree and/or Voronoi tessellation is left behind on each of the plates, one being a mirror image of the other. The mechanism is ubiquitous and reproducible in many viscous fluids. Manual separation at ambient pressure can result in the formation of a region of Voronoi tessellation inside the overall feature of a viscous tree.

Fig 3: The two mirror-images of a viscous tree produced by manual separation of plates sandwiching acrylic paint (Rembrandt, cadmium yellow deep 210/3). The diameter of the tree is 10 cm, the height of the branches of the order of one millimeter.

Fig 4: Secondary cavitation of sheets in the laboratory during manual separation of plates suspended by vinylpolysiloxan. In this experiment, the separation of the plates leads to the formation of a viscous tree, but secondary cavitation also takes place in foams.

**Figure 1a and 1b**

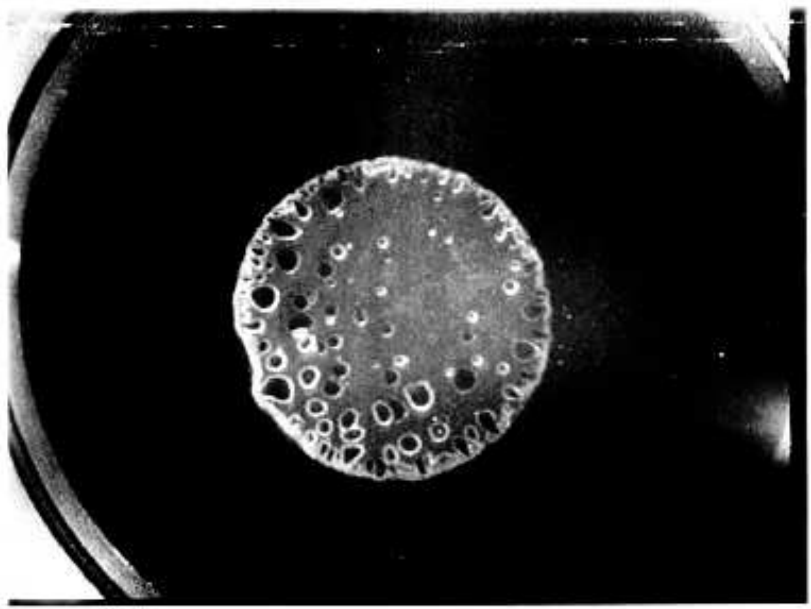

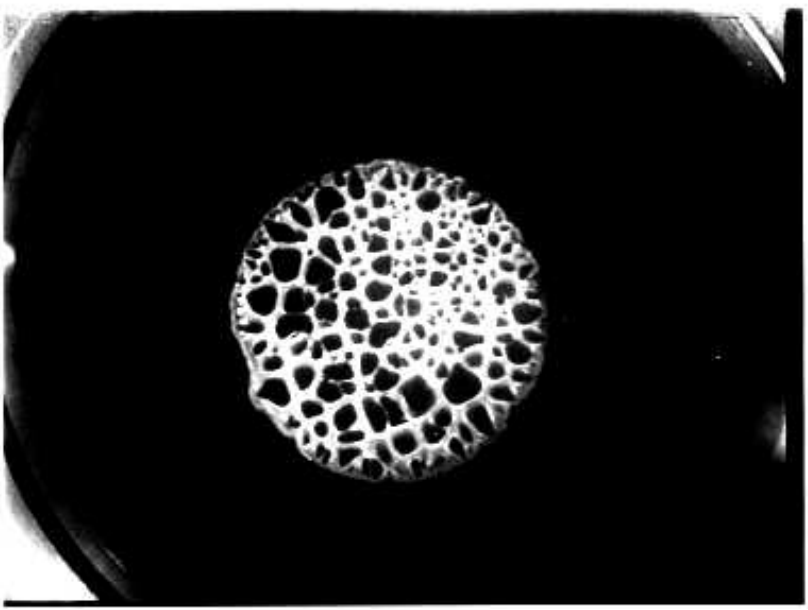

**Figure 2**

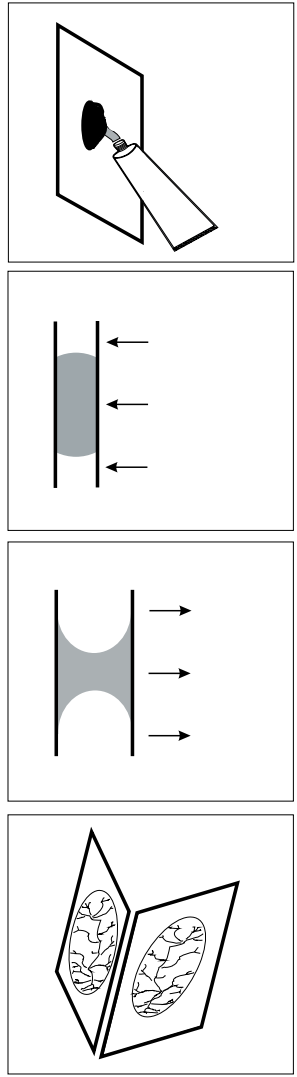

**Figure 3**

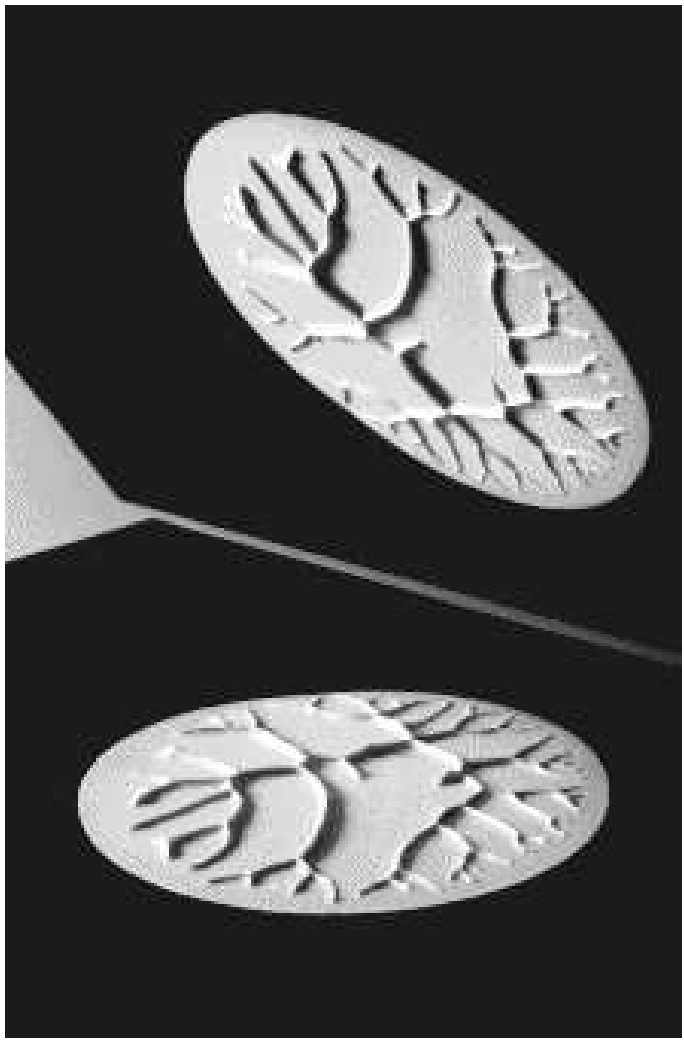

Figure 4

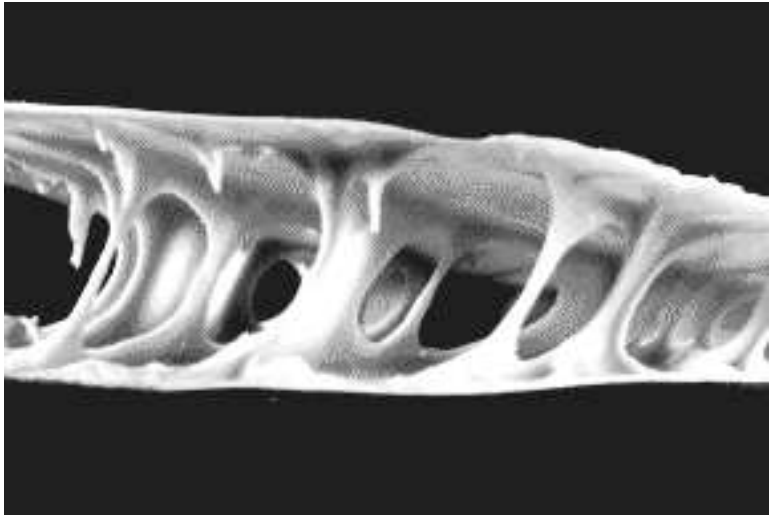